\documentclass[%
aip,
jmp,%
amsmath,amssymb,
reprint,%
]{revtex4-1}

\usepackage{graphicx}
\usepackage{bm}
\usepackage[T1]{fontenc}       
\usepackage{graphicx}
\usepackage[font=small,labelfont=bf,justification = justified]{caption}
\usepackage{titlesec}
\titlespacing*{\section}{0pt}{1.1\baselineskip}{\baselineskip}

\usepackage{float}
\usepackage{subcaption}
\usepackage{bigstrut}
\usepackage[table,xcdraw]{xcolor}
\usepackage[export]{adjustbox}
\usepackage{multirow}
\usepackage{epstopdf}
\DeclareGraphicsExtensions{.eps}

\usepackage{array}
\newcolumntype{M}[1]{>{\centering\arraybackslash}m{#1}}
\usepackage{tabularx}
\usepackage{booktabs}

\begin{document}

\preprint{AIP/123-QED}

\title[Visualizing the motion of graphene nanodrums]{Visualizing the motion of graphene nanodrums}

\author{Dejan Davidovikj}
\affiliation{ Kavli Institute of Nanoscience, Delft University of Technology, Lorentzweg 1, 2628 CJ Delft, The Netherlands}
\author{Jesse J Slim}
\affiliation{ Kavli Institute of Nanoscience, Delft University of Technology, Lorentzweg 1, 2628 CJ Delft, The Netherlands}
\author{Santiago J Cartamil-Bueno}
\affiliation{ Kavli Institute of Nanoscience, Delft University of Technology, Lorentzweg 1, 2628 CJ Delft, The Netherlands}
\author{Herre S J van der Zant}
\affiliation{ Kavli Institute of Nanoscience, Delft University of Technology, Lorentzweg 1, 2628 CJ Delft, The Netherlands}
\author{Peter G Steeneken}
\affiliation{ Kavli Institute of Nanoscience, Delft University of Technology, Lorentzweg 1, 2628 CJ Delft, The Netherlands}
\author{Warner J Venstra}
\affiliation{ Kavli Institute of Nanoscience, Delft University of Technology, Lorentzweg 1, 2628 CJ Delft, The Netherlands}
\affiliation{Quantified Air, Lorentzweg 1, 2628 CJ Delft, The Netherlands}
\date{\today}

\begin{abstract} 
Membranes of suspended two-dimensional materials show a large variability in mechanical properties, in part due to static and dynamic wrinkles. As a consequence, experiments typically show a multitude of nanomechanical resonance peaks, which makes an unambiguous identification of the vibrational modes difficult. Here, we probe the motion of graphene nanodrum resonators with spatial resolution using a phase-sensitive interferometer. By simultaneously visualizing the local phase and amplitude of the driven motion, we show that unexplained spectral features represent split degenerate modes. When taking these into account, the resonance frequencies up to the eighth vibrational mode agree with theory. The corresponding displacement profiles however, are remarkably different from theory, as small imperfections increasingly deform the nodal lines for the higher modes. The Brownian motion, which is used to calibrate the local displacement, exhibits a similar mode pattern. The experiments clarify the complicated dynamic behaviour of suspended two-dimensional materials, which is crucial for reproducible fabrication and applications.
\end{abstract}

\maketitle
\section{Introduction} 
\indent Nanomechanical devices from suspended graphene and other two-dimensional materials have received growing interest in the past few years~\cite{bunch07, castellanos13, wang15, cartamil15}, and their application in sensitive pressure, gas and mass sensors has been proposed~\cite{bunch08,smith13pressure,bunch12,bunch15,zettl08,sakhaee08mass, dolleman15}. Available techniques to study the mechanical properties of such membranes include quasi-static indentation~\cite{poot08,nicholl15} and dynamic response analyses in frequency- and time-domains~\cite{bunch07,vanderzande10arrays, bachtold11damping,castellanos13,leeuwen14,wang15}. These experimental studies show a large variability in the mechanical properties~\cite{vanderzande10arrays, barton11,castellanos13}, and to understand the intricate dynamic behaviour of suspended graphene, it is necessary to detect its motion with spatial resolution. While initial experiments were done on structures with intentionally broken radial symmetry~\cite{bachtold08, wang14}, the local phase of the membrane motion was not measured in these cases, which makes identification of the mode difficult, especially for the higher modes and in the presence of small imperfections. Moreover, previous experiments did not measure absolute displacements, which makes the acquisition of quantitative displacement profiles of the fundamental and higher modes impossible.\\ 
\indent Here, we visualize the motion of two-dimensional nanodrums with unprecedented resolution and sensitivity using a phase-sensitive scanning interferometer. The driven motion and the non-driven Brownian motion of a suspended few-layer graphene resonator vibrating at very high frequencies is detected up to the eighth vibrational mode. The phase information enables a reconstruction of the time-evolution of the displacement profile. In the radially symmetric nanodrum we observe splitting of multiple degenerate modes, as well as a distortion of the mode structure. By visualizing the Brownian motion, the displacement profiles are calibrated as to obtain spatially resolved displacement amplitudes. The spatially resolved measurements enable a detailed examination of the mode structure, and provide a useful tool in the efforts towards reproducible fabrication of suspended two-dimensional materials.
\section{Results} 
\subsection{Experimental setup} Circular graphene nanodrums are fabricated by transferring exfoliated few-layer graphene on top of silicon substrates pre-patterned with circular holes, as is described in the Methods section and in Supplementary Section 1. Figure 1a shows the graphene nanodrum of interest, with a diameter of $5\,\mathrm{\mu m}$ and a thickness of $5\,\mathrm{nm}$ as confirmed by Raman spectroscopy and atomic force microscopy (Fig.~1b). The flexural motion of the nanodrum is detected using an optical interferometer, which has been used previously in frequency- and time-domain studies of the nanomechanical properties of 2D-materials ~\cite{azak07,bunch07,lee13,castellanos13,nicholl15}. Figure 1c shows the setup and a schematic cross-section of the graphene nanodrum. The drum is probed by a Helium-Neon laser, and the intensity variations caused by the interfering reflections from the moving membrane and the fixed silicon substrate underneath are detected with a photodiode, as is described in more detail in Supplementary Section 2.\\
\begin{figure*}
	\centering
	\captionsetup{justification=justified}
	\includegraphics[width=\textwidth]{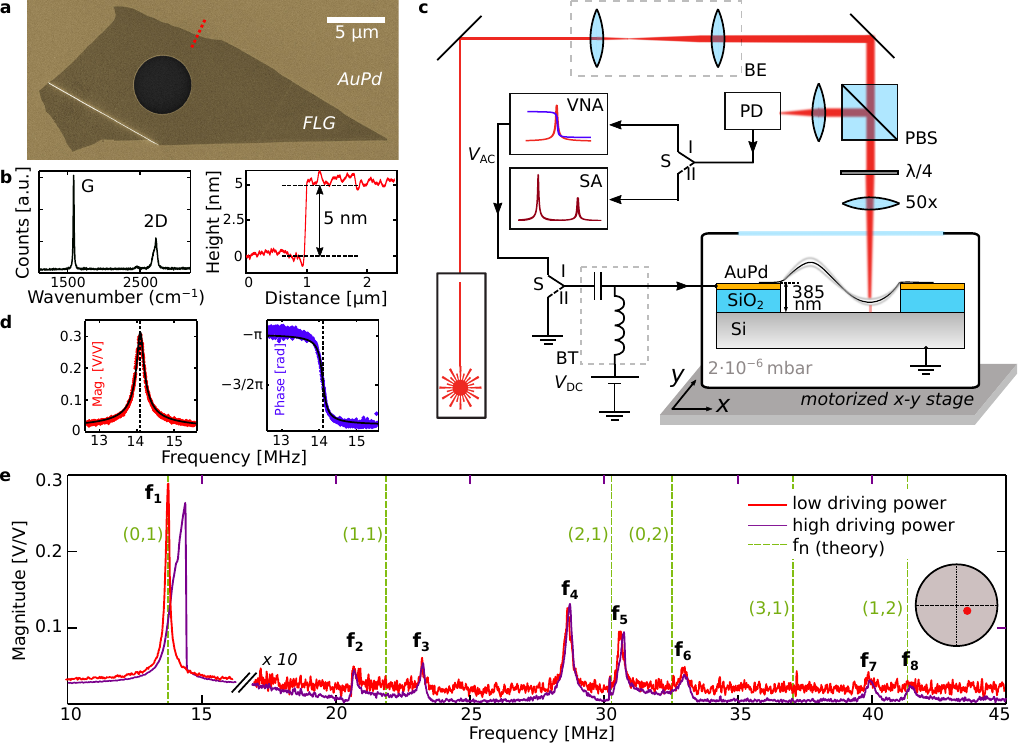}
	\caption[width=\textwidth]{\textbf{Scanning laser interferometry of graphene nanodrums.}\textbf{(a)} Scanning Electron Microscope (SEM) image of the graphene nanodrum. \textbf{(b)} Raman spectrum (left) taken at the centre of the drum; the relative height of the G and 2D peaks is characteristic of multi-layer graphene. Atomic Force Microscope (AFM) trace (right) taken along the red dashed line from (a), showing the flake thickness of 5 nm. \textbf{(c)}  Interferometric displacement detection is accomplished by focussing a HeNe laser beam ($\lambda = 632.8\,\mathrm{nm}$) on the nanodrum, while recording the interfering reflections from the graphene and the Si substrate underneath using a photodiode (PD). The sample is mounted on a motorized xy nanopositioning stage that scans the sample in a serpentine fashion, with a step size of 140 nm. BE: $3\mathrm{\times}$ Beam Expander; PBS: Polarized Beam Splitter. Two measurement types can be selected using switch $\mathrm{S}$:  $\mathrm{S\,=\,1}$ engages a phase-sensitive vector network analyser (VNA) measurement, while $\mathrm{S\,=\,2}$ is used to detect the Brownian motion of the nanodrum using a spectrum analyser (SA).  \textbf{(d)} VNA measurement (magnitude and phase) of the fundamental resonance mode, detected while probing at the centre of the drum (black curves: fitted response). \textbf{e)} VNA measurement showing the eight lowest resonance modes of the nanodrum, when driven at $2.2\,\mathrm{mV}$ (red) and $8.9\,\mathrm{mV}$ (black). Eight resonance peaks are detected, which are indexed 1-8 starting at the fundamental mode (for clarity, the magnitude of modes 2-8 is scaled $\mathrm{10\times}$).}

\end{figure*}

\indent The sample can be moved in-plane with the graphene (x-y) using a motorized nanopositioning stage. Compared to a scanning mirror, moving the sample does not affect the intensity of the incident light such that the transduction gain of the setup remains constant. This makes a calibration of the displacements possible, as will be shown below. With a step size of $140\,\mathrm{nm}$, the spectral response is measured at $1500$ points spatially distributed across the suspended part of the drum, which is sufficient to visualize the displacements associated with higher vibration modes, which exhibit an increasing number of nodal lines.\\
\indent Two measurement types can be selected by setting the switch $\mathrm{S}$ (see Fig.~1c). When $\mathrm{S\,=\,1}$, the complex response (magnitude and phase) to an electrostatic driving signal is measured using a vector network analyser (VNA). When $\mathrm{S\,=\,2}$, the driving signal is switched off, and the Brownian motion of the membrane is detected using a spectrum analyser (SA).  Figure 1d shows the magnitude and phase response ($\mathrm{S\,=\,1}$) at the fundamental resonance mode, which corresponds well to an harmonic oscillator. Figure 1e shows the response of the drum at higher driving frequencies, taken at low (red curve) and higher (black curve) driving voltages, when probing close to the centre of the drum. At strong driving, a multitude of peaks is detected, which are labelled $f_\mathrm{i}$, in accordance with their position in the spectrum. 

The resonance frequencies calculated for a perfect circular membrane are also displayed~\cite{mn}. The measured resonance frequencies are conspicuously different from the calculated ones, which raises the central question addressed in this work: Which mode indices correspond to each of the observed resonance peaks?

\subsection{Visualizing driven motion} The motion of the graphene nanodrum is made visible using a phase-sensitive scanning interferometer. To demonstrate the technique, we set out by measuring the frequency response of mode 3 at five different positions on the drum, as illustrated in Fig.~2a. The amplitude and phase responses are fit to a harmonic oscillator function, and the fits are shown in Fig.~2b. 
\begin{figure*}[]
	\includegraphics[width=\textwidth]{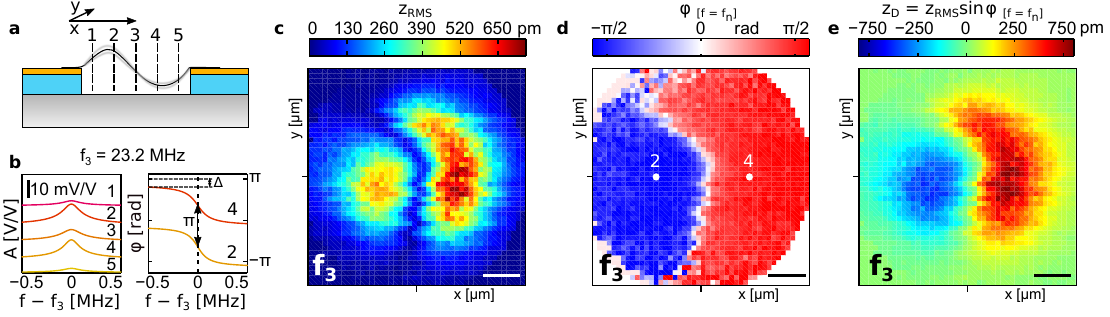}
	\caption[width=\textwidth]{\textbf{Spatially-resolved measurements.}\textbf{(a)} Individual frequency response lines taken at positions 1-5 using the VNA ($\mathrm{S\,=\,1}$), while applying a driving signal at $f_\mathrm{3}$. \textbf{(b)} Recorded magnitude (left) and phase (right) response at $f_\mathrm{3}$. A phase difference $\mathrm{\pi}$ indicates that on these locations the drum moves in opposite direction. \textbf{c)} Root-mean-square displacement $z_\mathrm{RMS}$ of mode 3, taken at a step size (x,y) of $140 \,\mathrm{nm}$. The diffraction-limited spot size of the probe laser is $1.3\,\mathrm{\mu m}$, which causes some loss of spatial resolution. \textbf{(d)} Local phase response, $\phi_\mathrm{R}$, of the nanodrum, showing that two halves of the drum move in opposite directions. \textbf{(e)} Reconstructed displacement field map as obtained by  $z_\mathrm{D}=z_\mathrm{RMS} \sin \phi$. Scale bars in (c-e): $1\,\mathrm{\mu m}$.}
\end{figure*}
From the responses it is observed that, while two halves of the drum move at a comparable amplitude, their phase differs by $\mathrm{\pi}$. This indicates that at positions 2 and 4 the graphene moves in opposite direction, as is the case for a $\mathrm{(1,1)}$ mode~\cite{mn}. Following this procedure, a more refined measurement is performed. Figure 2c,d display on a color scale the fitted peak height, $z_\mathrm{RMS}$, and phase responses $\mathrm{\phi}$ at the resonance peak frequency, measured on a square grid with a spacing of 140 nm. The amplitude response reveals two anti-nodal points which are separated by a nodal line, and the phase response shows that on either side of the nodal line the graphene moves in opposite direction. While for the $\mathrm{(1,1)}$ mode this phase behaviour appears trivial, we will show below that the phase information is a requisite to understand the motion of higher modes. Figure 2e shows a snapshot colormap of the membrane movement, as obtained by $z_\mathrm{D} = z_\mathrm{RMS}\sin\phi$. A time-lapsed visualization of the motion over the oscillation period is provided in the Supplementary Video.\\
\indent In a similar way, the motion of the other resonance peaks of Fig.~1e is visualized. Figure 3 a-h show the mode shapes that correspond to $f_\mathrm{1} - f_\mathrm{8}$, together with the theoretical shapes, which were obtained by finite-element calculations for a circular membrane. The fundamental mode was probed at a reduced driving voltage, as to maintain a linear response as in Fig.~1e, red curve. The measurements show unambiguously that the peaks observed in the spectrum of Fig.~1e are the result of split degenerate modes $\mathrm{(1,1)}$, $\mathrm{(1,2)}$, and $\mathrm{(2,1)}$. The displacement profiles of modes 1-4 (Fig.~3a-d) are in reasonable agreement with the theoretically calculated mode shapes. Other modes however, in particular the ones with higher indices (i.e., 5, 6 and 8 as shown in Fig.~3e-g) show a large discrepancy. Clearly, an imperfection is present whose influence on the location of the nodal lines grows with the mode index. Note that this imperfection is not visible in the SEM image of Fig.~1a, but has a large impact on the mode shapes. Table 1 summarizes the experimental and theoretical resonance frequencies and mode shapes, and will be described further in the next section.
\begin{table}
	\centering
	\caption{Experimental and theoretical resonance frequencies of the graphene nanodrum}
	\begin{tabular}{|r|c|c|c|c|M{1cm}|}
		\cline{2-6}    \multicolumn{1}{r|}{} & \multicolumn{2}{c|}{experiment} & \multicolumn{3}{c|}{theory} \bigstrut\\
		\hline
		\multicolumn{1}{|c|}{$\mathrm{i}$} & $f_\mathrm{i}\,\mathrm{[MHz]}$ & $f_\mathrm{i}/f_\mathrm{0}$ & $f_\mathrm{i}/f_\mathrm{0}$ & \multicolumn{1}{M{1.5cm}|}{$\mathrm{(m,n)}$} & \multicolumn{1}{M{1.5cm}|}{shape} \bigstrut\\
		\hline
		\multicolumn{1}{|c|}{$1$} & $14.1$  & $1$     & $1$     & $\mathrm{(0,1)}$ &  \multicolumn{1}{c|}{\includegraphics[valign = m,scale=0.12]{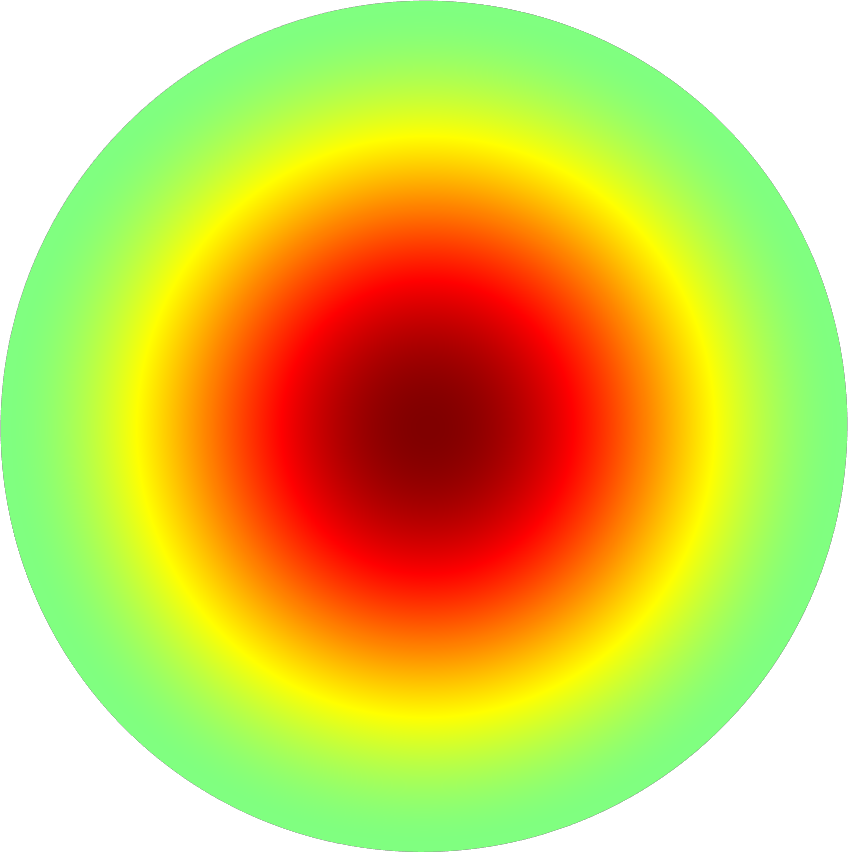}}\\
		\hline
		\multicolumn{1}{|c|}{$2$} & $20.69$ & $1.467$ & \multirow{2}[4]{*}{$1.593$} & \multirow{2}[4]{*}{$\mathrm{(1,1)}$} & \multicolumn{1}{c|}{\multirow{2}[4]{*}{\includegraphics[valign = m,scale=0.09]{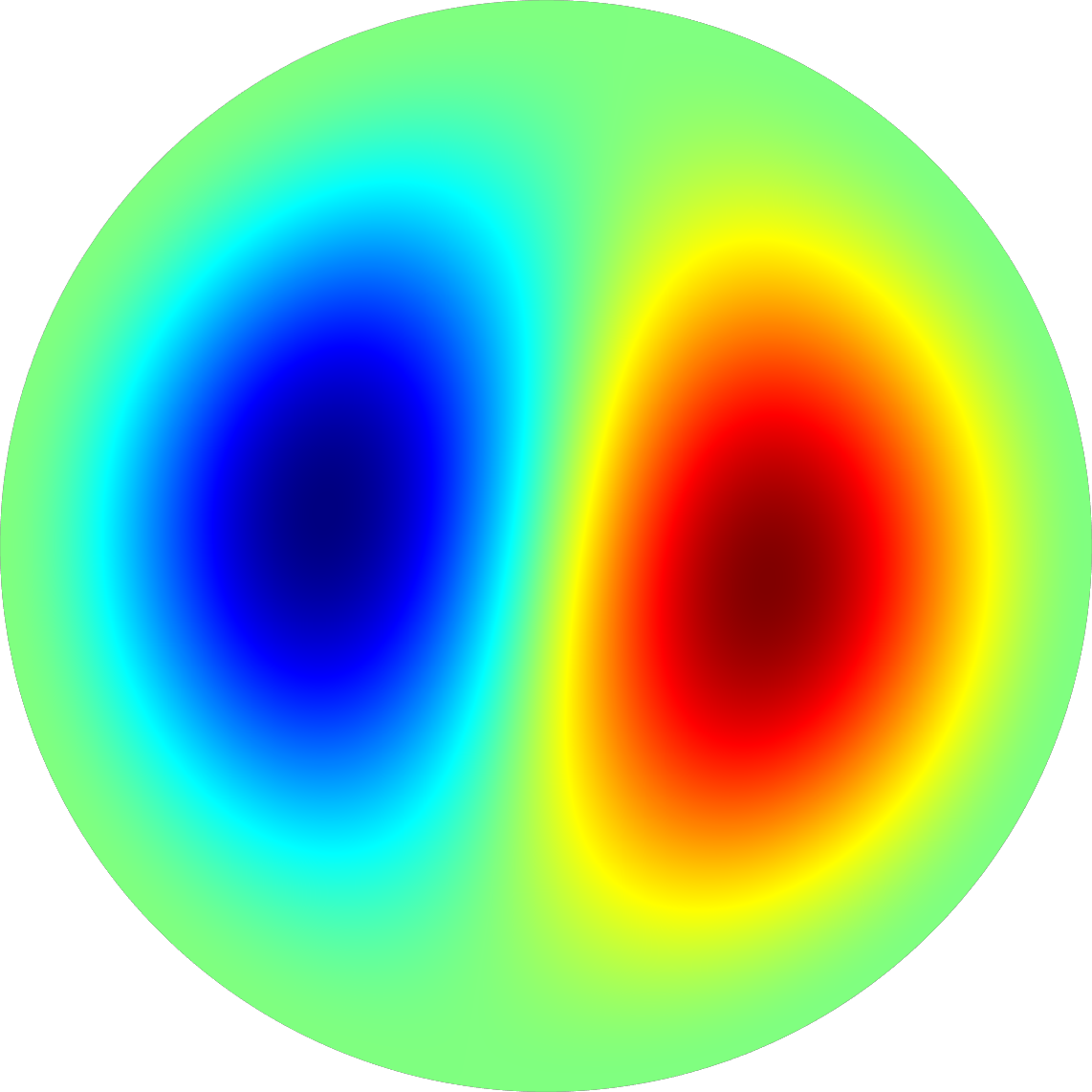}}} \bigstrut\\
		\cline{1-3}    \multicolumn{1}{|c|}{$3$} & $23.24$ & $1.648$ &       &       & \multicolumn{1}{c|}{} \bigstrut\\
		\hline
		\multicolumn{1}{|c|}{$4$} &$28.73$ & $2.038$ & \multirow{2}[4]{*}{$2.135$} & \multirow{2}[4]{*}{$\mathrm{(2,1)}$} & \multicolumn{1}{c|}{\multirow{2}[4]{*}{\includegraphics[valign = m,scale=0.09]{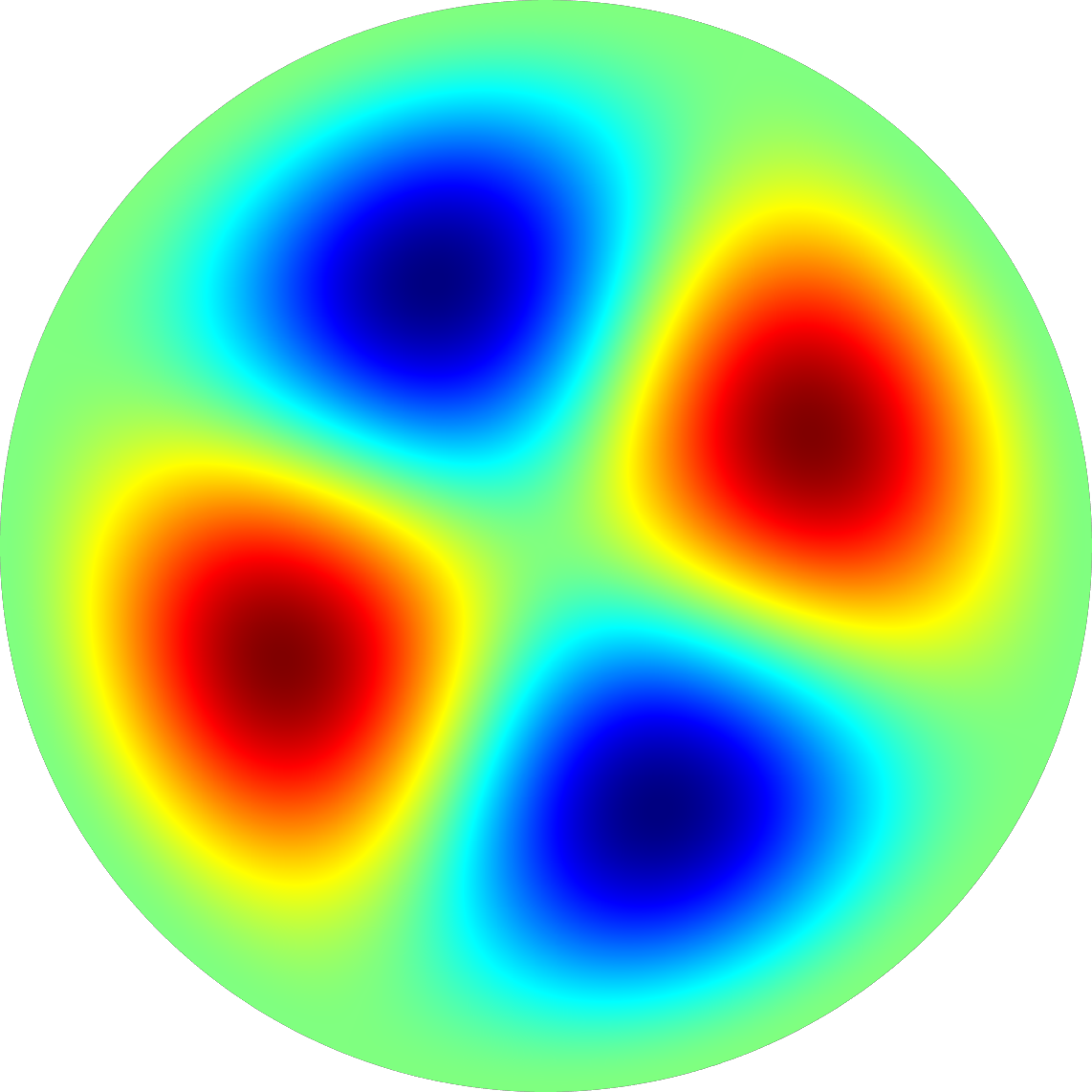}}} \bigstrut\\
		\cline{1-3}    \multicolumn{1}{|c|}{$5$} & $30.75$ & $2.181$ &       &       & \multicolumn{1}{c|}{} \bigstrut\\
		\hline
		\multicolumn{1}{|c|}{$6$} & $33$    & $2.340$ & $2.295$ & $\mathrm{(0,2)}$ &  \multicolumn{1}{c|}{\includegraphics[valign = m,scale=0.12]{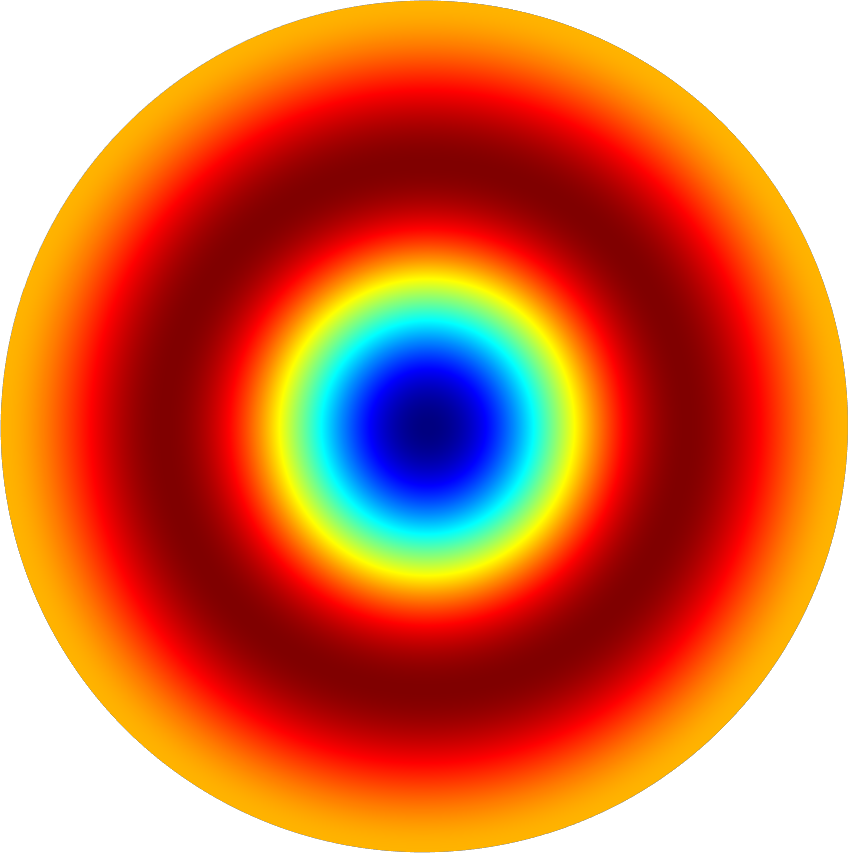}}\\
		\hline
		\multicolumn{1}{|c|}{-} & -     & -     & \multirow{2}[4]{*}{$2.653$} & \multirow{2}[4]{*}{$\mathrm{(3,1)}$} & \multicolumn{1}{c|}{\multirow{2}[4]{*}{\includegraphics[valign = m,scale=0.12]{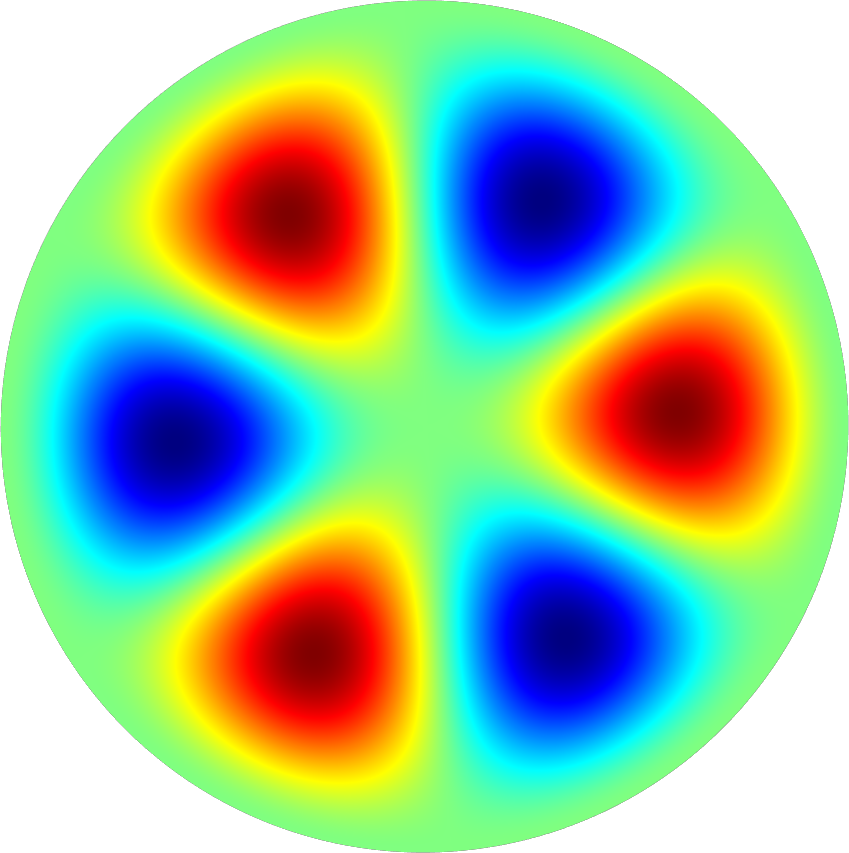}}} \bigstrut\\
		\cline{1-3}    \multicolumn{1}{|c|}{-} & -     & -     &       &       & \multicolumn{1}{c|}{} \bigstrut\\
		\hline
		\multicolumn{1}{|c|}{$7$} & $39.93$ & $2.832$ & \multirow{2}[3]{*}{$2.917$} & \multirow{2}[3]{*}{$\mathrm{(1,2)}$} & \multicolumn{1}{c|}{\multirow{2}[3]{*}{\includegraphics[valign = m,scale=0.12]{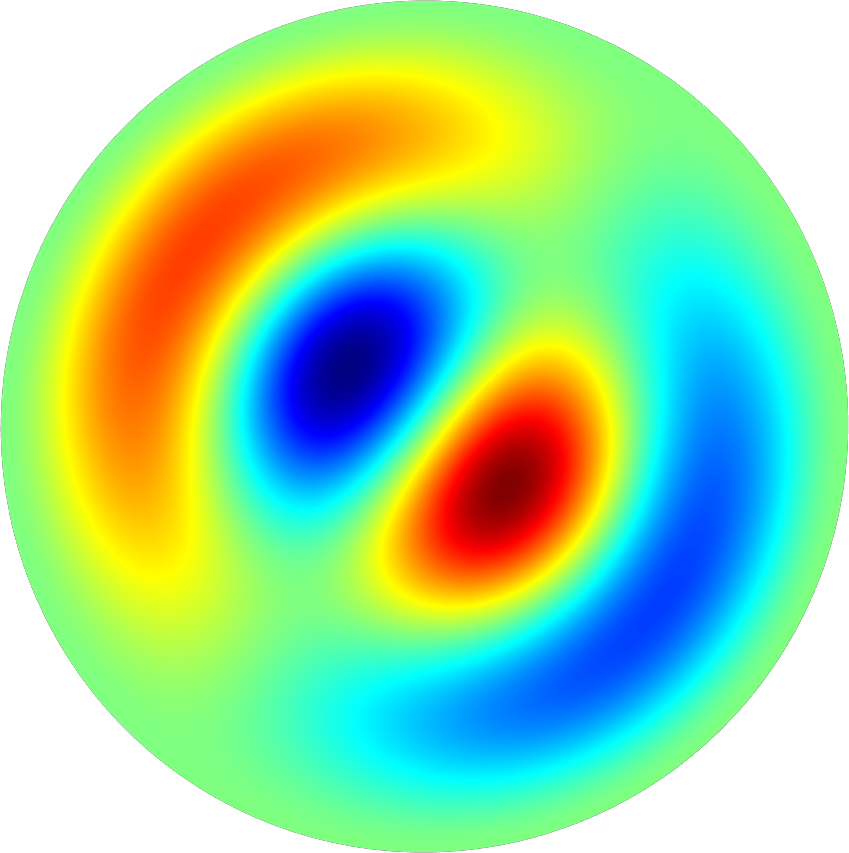}}} \bigstrut\\
		\cline{1-3}    \multicolumn{1}{|c|}{$8$} & $41.41$ & $2.937$ &       &       & \multicolumn{1}{c|}{} \bigstrut[t]\\
		\hline
	\end{tabular}%
	\label{tab:addlabel}%
\end{table}%

\begin{figure*}
	\includegraphics[width=\textwidth]{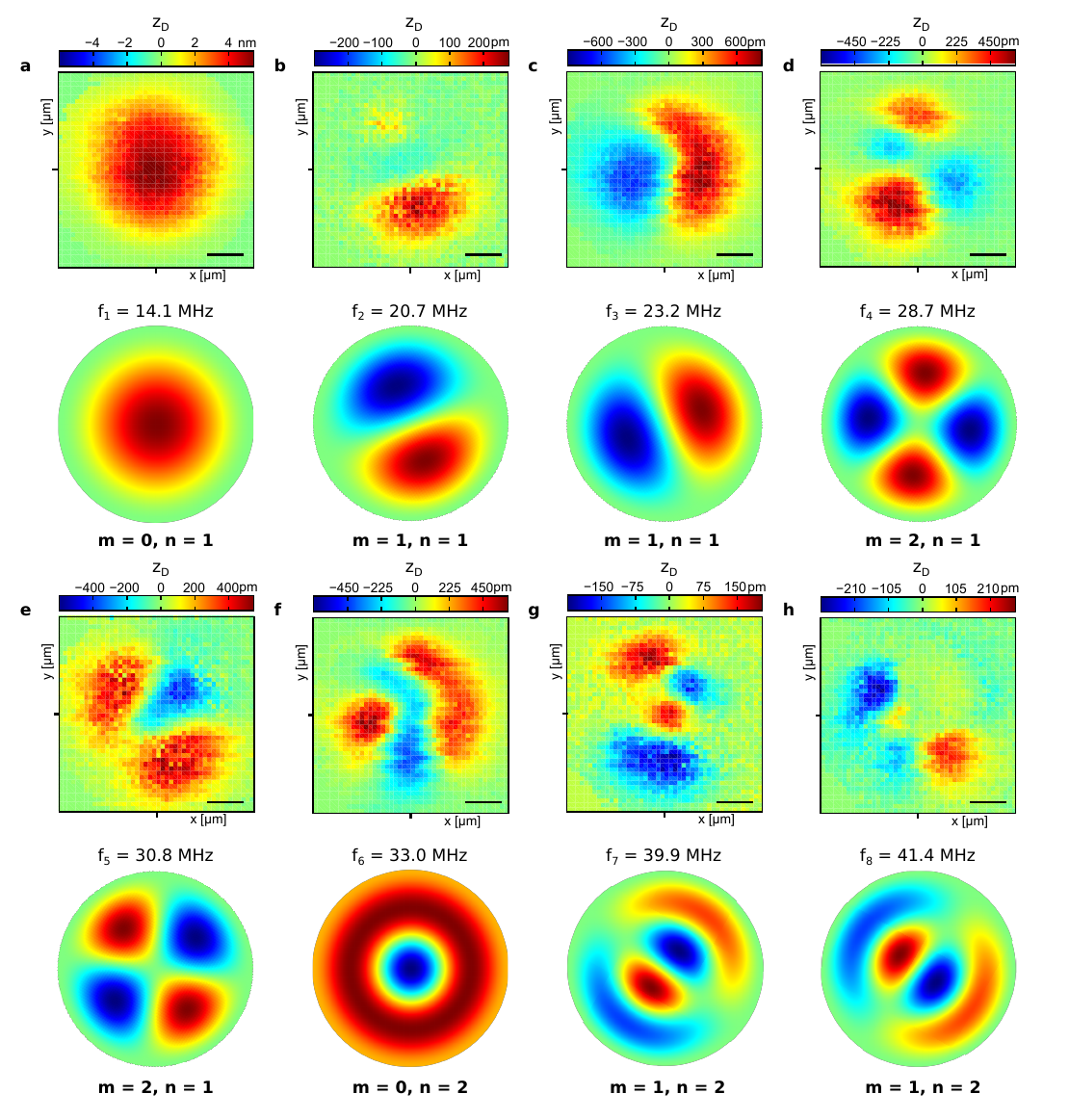}
	\caption[width=\textwidth]{\textbf{Visualizing resonant motion.} \textbf{(a-h)} Top: experimental data; bottom: finite-element calculation. The modes predicted by the calculation are indexed by $\mathrm{(m,n)}$. Panels (b) and (c) show that the nanodrum hosts a split degenerate $\mathrm{(1,1)}$ mode, while also the $\mathrm{(2,1)}$ mode is split, as is shown in panels (d) and (e). The displacement profile measured in (f) resembles a $\mathrm{(0,2)}$ mode, which is distorted due to an imperfection as will be discussed in the main text. (g) and (h) reveal a degenerate $\mathrm{(1,2)}$mode. Scale bars: $1\,\mathrm{\mu m}$.}
\end{figure*}

\subsection{Visualizing Brownian motion} The above experiments illustrate in detail the mode structure of a driven graphene nanodrum, and it is interesting to compare these driven measurements with the displacements that are the result of thermal fluctuations. Compared to silicon carbide micro-disk resonators, whose thermal motion was studied recently~\cite{wang14sic}, graphene nanodrums have a very low reflectivity and a 10-100 times lower mechanical Q-factor. Nevertheless, the present technique is sensitive enough to visualize their Brownian motion.\\
\indent To study the Brownian motion, the switch is set to $\mathrm{S\,=\,2}$, in order to switch-off the driving signal and to the record displacements with a spectrum analyser. Figure 4a shows an example of a thermal noise spectrum, taken close to the centre of the drum. Three vibrational modes are observed that resemble the lowest three resonances of Fig.~1e, albeit at somewhat lower frequencies. The difference in frequency results from the absence of the electrostatic force: as $V_\mathrm{dc}=0$, no force is exerted on the graphene drum. Compared to teh driven measurement, a part of the mechanical tension is released, which causes the resonance frequencies to tune to a lower value. Supplementary Section 3 discusses the tuning of the resonance frequency as well as the optimization of the signal-to-noise ratio by adjusting $V_\mathrm{dc}$. With the same step size as in the driven measurement, we map the first three mode shapes and plot the thermal RMS displacement as a function of position. Figure 4b shows the Brownian motion of the fundamental $\mathrm{(0,1)}$ mode, and 4c and 4d show a splitting of the degenerate $\mathrm{(1,1)}$ mode, in close agreement with the amplitude map of the driven motion. Note that in the absence of a driving signal the phase is not measured as it diffuses within the measurement integration time.\\

\begin{figure*}
	\includegraphics[width=\textwidth]{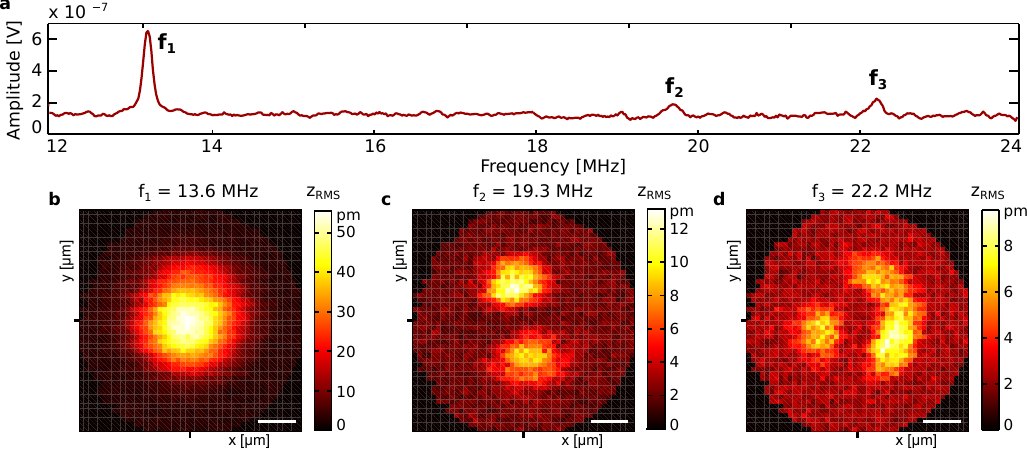}
	\caption[width=\textwidth]{\textbf{Visualizing Brownian motion.} (a) Spectrum analyser measurements ($\mathrm{S\,=\,2}$) taken without applying a driving signal ($V_\mathrm{ac} = 0$) reveal three vibrational modes. (b-d) Spatial maps of the RMS Brownian displacements at each of the vibrational modes. The Brownian mode shapes  correspond well with the ones observed when the drum is resonantly driven (Fig.~3(a-c)). The slightly lower frequencies are the result of the absence of a dc-voltage $V_\mathrm{dc} = 0\,\mathrm{V}$, which results in a sightly lower mechanical tension in the drum, as is discussed in Supplementary Section 1.}
\end{figure*}

\indent For applications of suspended two-dimensional materials, it is important to quantify the displacements associated with the motion. The RMS displacement of a nanomechanical resonator can be obtained by measuring its Brownian motion~\cite{Li07}. From the power spectral density of the signal measured at the centre of the drum, $S_\mathrm{VV}(f) = S_\mathrm{VV}^\mathrm{w} + \alpha S_\mathrm{zz}(f)$, the noise floor $S_\mathrm{VV}^\mathrm{w}$ and the transduction factor $\alpha\,\mathrm{[V^\mathrm{2}/m^\mathrm{2}]}$ are calculated. The thermal displacement noise spectral density of the fundamental mode at the drum centre is given by $S_\mathrm{zz}(f) = \frac{k_\mathrm{B}Tf_\mathrm{1}}{2\pi ^3 m_\mathrm{eff,1}Q_\mathrm{1}[(f^\mathrm{2} - f_\mathrm{1}^\mathrm{2})\mathrm{^2} + (ff_\mathrm{1}/Q_\mathrm{1})\mathrm{^2}]}$, where $f_\mathrm{1}$ is the resonance frequency, $Q_\mathrm{1}$ is the quality factor and $m_\mathrm{eff,1} = 0.2695\, m_\mathrm{total}$ its effective mass~\cite{hauer13}. With an incident optical power of $0.8\,\mathrm{mW}$, the noise floor equals $\mathrm{11\,fm/\sqrt{Hz}}$, which enables the detection of the three resonance modes. The transduction factor equals $\alpha = 3.75 \times 10^\mathrm{11}\, \mathrm{V^\mathrm{2}/m^\mathrm{2}}$, and using this number, all detected displacement signals in the experiments of Fig.~3 and Fig.~4 are converted to absolute displacements, and indicated in the respective colour bars.

\section{Discussion} 
From the summary of the measurement results presented in Table 1, it becomes clear that the ratios of the higher harmonics to the fundamental mode, $\mathrm{f_\mathrm{n} / f_\mathrm{0}}$, deviate from the theoretically expected frequencies for a membrane resonator~\cite{mode31}. Deviations range from 0.7 \% for mode $f_\mathrm{8}$ to 8.6 \% for mode $f_\mathrm{2}$. While the spatial maps show that the difference between the measured and theoretical mode shapes increases with the mode index, there is no obvious correlation between the differences in the resonance frequencies and the distortion of the mode shapes. For example, the mode shape of $f_\mathrm{3}$ is in good agreement, while the mode shape of $f_\mathrm{8}$ bears almost no resemblance to the theoretical calculation. Interestingly, comparing them in the frequency domain, $f_\mathrm{8}$ differs by only 0.7 \%, and $f_\mathrm{3}$ by 3.3 \% from its theoretical value. The same holds for the radially symmetric mode $\mathrm{(0,2)}$, whose frequency is within 2 \% of the calculation, while its mode shape is highly distorted. Thus, the lower modes appear more robust against imperfections, possibly due to the lower number of nodal lines -- a tendency that is confirmed by finite-element simulations provided in Supplementary Section 4. These findings are of particular interest in the light of the recently proposed nanomechanical schemes to detect the \emph{geometry} of adsorbed masses~\cite{dohn05,hanay15}, which rely on an accurate description of the mode shapes. In such schemes, the splitting of the degenerate modes, which is also observed in the Brownian motion and emerges from the structural imperfections in our experiments, could be used to provide information about the geometry of the adsorbed mass.\\
\indent It is interesting to further investigate the origins of the mode-splitting and the progressive distortion of the mode structure for the higher modes. To this end, we map the local stiffness of the drum using peak-force AFM (measurement details provided in Supplementary Section 4). The analysis reveals local inhomogeneities in the membrane that went unnoticed during optical and electron microscopy inspection. These may be the result of a uniaxial residual tension in the drum, introduced while 'peeling-off' the graphene flake during the transfer to the substrate~\cite{castellanos14}. Similar effects were observed in all studied drums, and it can thus be expected from this work that these are inherent to suspended two-dimensional materials fabricated by exfoliation and dry transfer. A finite-element calculation that takes this feature into account results in a better agreement between the predicted mode shapes and the measurements, as is discussed in Supplementary Section 4.\\
\indent Besides the displacements on resonance described above, other parameters can be visualized with spatial resolution. For example, the local resonance frequency, the Q-factor and the noise floor give a wealth of information about the device and the detector, such as the local temperature distribution in the device, and the local reflectivity of the substrate. These examples are discussed in more detail in Supplementary Section 5. Spatially resolved measurements are a valuable tool to analyse the dynamic properties of two-dimensional materials, and may be used to address open questions such as the origin of their low mechanical Q-factors~\cite{helgee14,kramer15}, as well as to assess fabrication quality and reproducibility. These are essential in order to exploit opportunities that arise in new applications as hybrid nano-electromechanical systems, that fuse excellent mechanical properties with exotic traits such as a negative thermal expansion coefficient and Poisson's ratio~\cite{yoon11negativeG,jiang14bPpoisson} and electromechanical~\cite{duerloo13,wu14} and optoelectronic couplings~\cite{hone10MoS2}.\\
\indent In conclusion, we visualize the motion of micrometer-scale graphene drums vibrating at very high frequencies with a lateral resolution of $\mathrm{140\,nm}$ and a displacement resolution of $\mathrm{11\,fm/\sqrt{(Hz)}}$. The driven and non-driven thermal displacement profiles of the radially symmetric drum reveal the motion associated with nanomechanical resonance peaks up to the eighth vibrational mode. The spatial technique presented in this work complements the frequency- and time-domain techniques presently available, and is crucial to obtain a complete description of the dynamic behaviour of suspended two-dimensional materials.

\section*{Methods}
\subsection{Fabrication of graphene nanodrums} The nanodrums were fabricated on a p-type silicon wafer with a $285\,\mathrm{nm}$ thick layer of thermal silicon oxide. First the top electrodes, the circular cavities and the bonding pads are patterned using negative resist and e-beam lithography. A layer of Ti/AuPd ($5/95\,\mathrm{nm}$) is evaporated on top, to provide a smooth and electrically conducting surface for the adhesion of graphene. By a lift-off of the AuPd, metallic islands are fabricated, which serve as a hard mask during the subsequent reactive ion etching of $\mathrm{SiO_2}$. The thickness of the metallization was chosen to obtain a cavity depth of $385\,\mathrm{nm}$, which optimizes the responsivity of interferometric measurements at the wavelength of the probing laser ($\lambda = 632.8\,\mathrm{nm}$), as is described in detail in Supplementary Section 2. In the final step, few-layer graphene flakes are mechanically exfoliated from natural crystals and deposited on top of the substrates using a dry transfer method~\cite{castellanos14}.

\subsection{Electrostatic actuation} The suspended graphene drum is electrostatically driven by applying a voltage, $V_\mathrm{dc}+V_\mathrm{ac}$ to the AuPd pad, while connecting the Si back-gate to ground. The dc-voltage tunes the static tension in the graphene flake, while the ac-voltage excites its resonant motion. The measurements of Fig.~1e are carried out with $V_\mathrm{ac} = 2.2\,\mathrm{mV}$ (red curve, linear response) and $V_\mathrm{ac} = 8.9\,\mathrm{mV}$ (black curve, nonlinear response). In both cases a dc-bias voltage $V_\mathrm{dc} = 3\,\mathrm{V}$ is used to amplify the time-dependent actuation force via $F(t)\propto V_\mathrm{dc}V_\mathrm{ac}\sin(2\pi f t)$, and to electrostatically control the tension as to optimize the signal-to-noise ratio, as is described in detail in Supplementary Section 4.  The xy-stage is actuated using NewFocus type 8301 picomotors with a type 8732 multi-axis driver. All experiments are conducted at room temperature, at a pressure of $\approx 2\times 10^{-6}\,\mathrm{mbar}$.

\section*{Acknowledgements}
We acknowledge discussions with Johan Dubbeldam. This work was supported by the Netherlands Organisation for Scientific Research (NWO/OCW) as part of the Frontiers of Nanoscience program, European Union Seventh Framework Programme under grant agreement $\mathrm{n{\circ}~604391}$ Graphene Flagship and the European Union's Seventh Framework Programme (FP7) under Grant Agreement $\mathrm{n{\circ}~318287}$, project LANDAUER.\\\\
\textbf{Competing Financial Interests} The authors declare no competing financial interests.\\
\textbf{Correspondence} Correspondence and requests for materials should be addressed to D.D. ~(email:d.davidovikj@tudelft.nl) or to W.J.V.~(email: w.j.venstra@tudelft.nl).

\newpage

\section*{Supplementary Information}
\setcounter{figure}{0}  
\subsection*{1.	Sample fabrication}
The samples are fabricated on a p-type <110> silicon wafer with a 285 nm thick layer of thermally grown silicon oxide. First, the resist is patterned using e-beam lithography. Next, layers of titanium (Ti, 5 nm) and gold-palladium (AuPd, 95 nm) are evaporated and a lift-off is performed in hot acetone. The resulting AuPd islands, shown in Fig. 1a, serve as a hard mask during the reactive ion etching of SiO2 (50 sccm CHF3 and 2.5 sccm Ar). This step defines the cavities (Figure S1b), and as the etching stops at the silicon, a flat reflective surface is obtained that forms the fixed mirror of the interferometer. The chips are then cleaned, and graphene flakes are transferred using an all-dry transfer method, which results in the device shown in Figure S1c.

\begin{figure}[h]
	\includegraphics[width=0.5\textwidth]{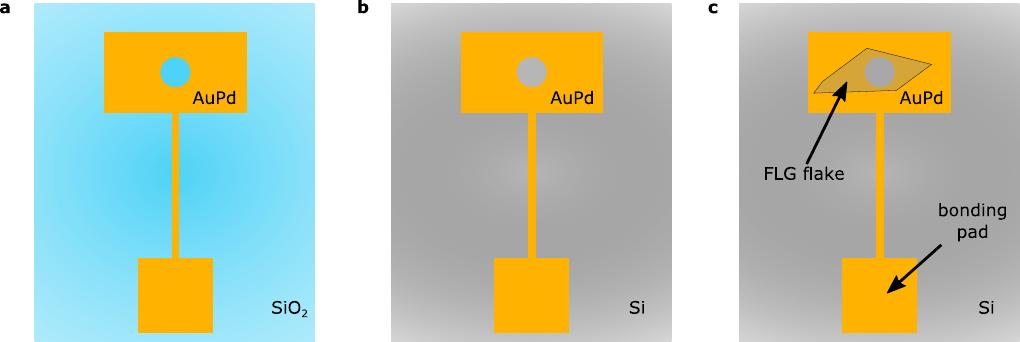}
	\caption[width=0.5\textwidth,justification = justified]{\textbf{Sample fabrication.} (a) Islands of titanium (5 nm) gold-palladium (95 nm) after e-beam patterning and lift-off. (b) The AuPd is used as a hard mask to etch the exposed $\mathrm{SiO_2}$ using Reactive Ion Etching (RIE): $\mathrm{CHF_3}$ (50 sccm) + Ar (2.3 sccm). (c) Graphene (or another 2-D material) is exfoliated and stamped on top of the 5 $\mathrm{\mu m}$ hole, defined by the metallic islands. Electrical contact to the nanodrum is established through the bonding pad.}
\end{figure}

\subsection*{2.	Optimization of the cavity depth}

The cavity length is adjustable by varying the thickness of the AuPd layer, and it is chosen to  maximize the responsivity at the laser wavelength for a range of graphene thicknesses as follows.  The reflectivity of the device, which consists of the three interfaces schematically depicted in Figure S2b, is calculated by [1]:\\

$\mathrm{R = |\frac{r_1+r_2e^{-i\delta_1}+r_3e^{-i\delta_2}+r_1r_2r_3e^{-i(\delta_1+\delta_2)}}{1+r_1r_2e^{-i\delta_2}+r_1r_3e^{-i(\delta_1+\delta_2)}+r_2r_3e^{-i\delta_2}}|^2}$
\\

Here, $\mathrm{\delta_1}$ and $\mathrm{\delta_2}$ represent the acquired phase while traveling through the different media indexed in Figure S2a.  Then $\mathrm{\delta_1 = \frac{2\pi n_1 N t_0}{\lambda}}$ with $\mathrm{N}$ being the number of layers and $\mathrm{t_0}$ the thickness of a single layer, accounts for the graphene, while $\mathrm{\delta_2 = \frac{2\pi n_0 N z_0}{\lambda}}$, with $\mathrm{z_0}$ being the rest position of the graphene, accounts for the cavity. The responsivity of the device, $\mathrm{\frac{dR}{dz}}$ , is plotted as a function of the cavity depth $\mathrm{\delta_1}$ , and graphene thickness in Figure S2b.  A cavity length of 385 nm optimizes the responsivity for flakes with thicknesses in a range between 1 - 30 layers (0.33 - 10 nm).

\begin{figure}[h]
	\includegraphics[width=0.5\textwidth]{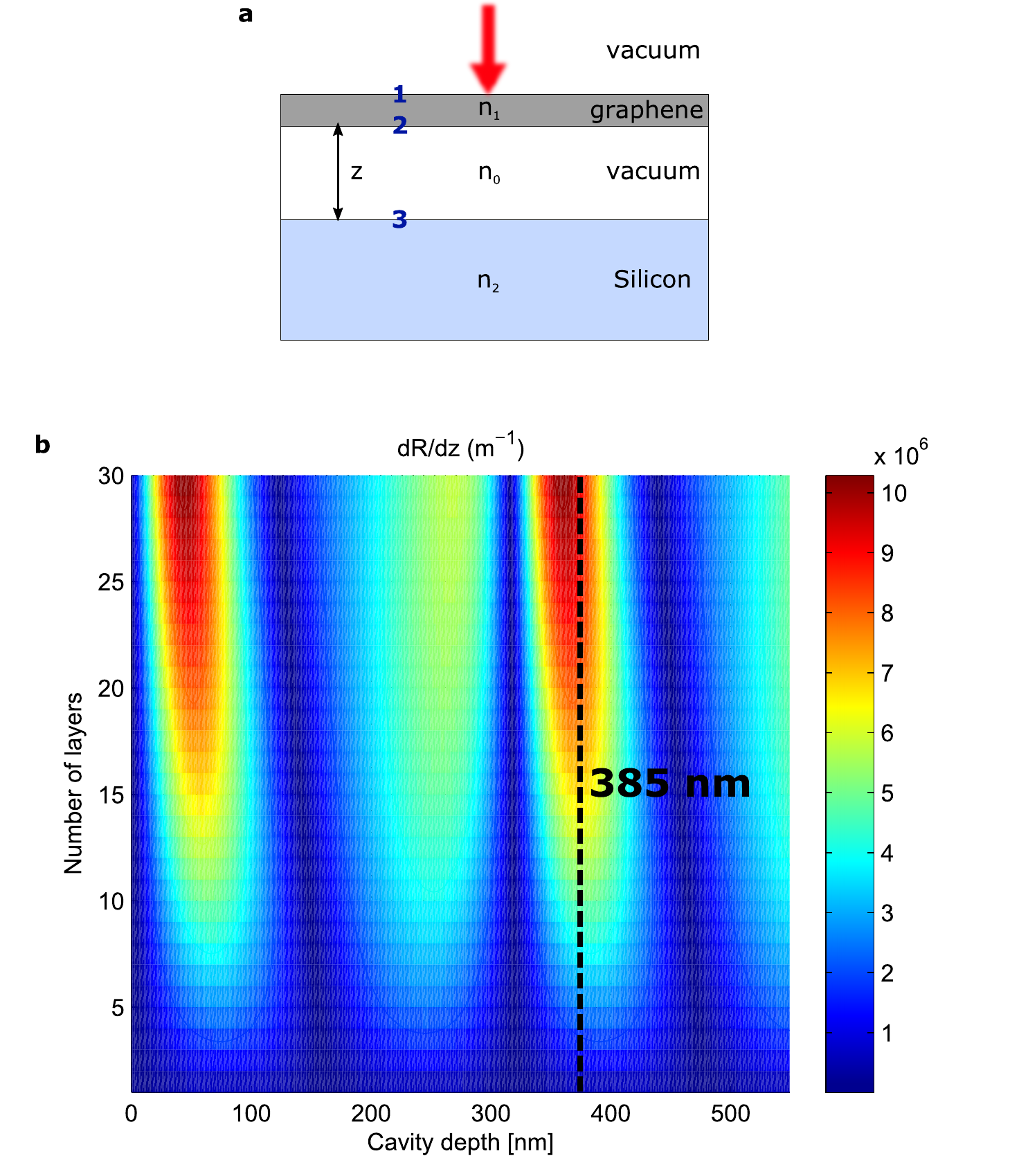}
	\caption[width=0.5\textwidth,justification = justified]{\textbf{Calculation of the responsivity.} (a) Schematic of the different media and interfaces. $\mathrm{n_i}$ is the refractive index of medium i (i=0,1,2); z denotes the cavity depth. The red arrow denotes the incident laser light. (b) Calculated responsivity as a function of cavity depth and graphene thickness. The black dashed line indicates the thickness chosen in order to optimize the responsivity for graphene thicknesses in the range between 1-30 layers. }
\end{figure}

\subsection{3.	Optimizing the signal by electrostatic tuning}

The effective spring constant of the nanodrum is adjustable by applying an electric field across the drum and the silicon substrate. Figure S3a shows spectra of the Brownian motion around the fundamental resonance mode as a function of the applied dc-voltage, $\mathrm{V_{dc}}$. Two regimes are present: for $\mathrm{|V_{dc}|}$< 1.5 V a spring weakening is observed, which is due to the attractive electrostatic force. Larger voltages result in a significant stretching of the membrane and the increased mechanical tension gives rise to a spring stiffening behaviour. The inset of Figure S3a shows the root-mean-square amplitude of the Brownian motion as a function of applied voltage. The amplitude and the signal-to-noise ratio (SNR) are maximized close to $\mathrm{V_{dc}}$ = 0, and the Brownian motion measurements discussed in the main text are performed at this setting.
In the driven measurements, an alternating voltage $\mathrm{V_{ac}}$  is applied, and in this case, applying a dc-voltage amplifies the electrostatic driving force via $\mathrm{F_{ac} \propto V_{ac}V_{dc}sin(2\pi ft)}$. The amplification counteracts the reduction of the SNR due to the spring stiffening in the case $\mathrm{V_{ac}}$ = 0. Figure S3b shows the dc-voltage dependence of the resonance frequency for the driven motion. In this case, the optimum SNR is achieved at $\mathrm{V_{dc}}$ = 3 V, and this setting was chosen for the the driven measurements.

\begin{figure}[h]
	\includegraphics[width=0.5\textwidth]{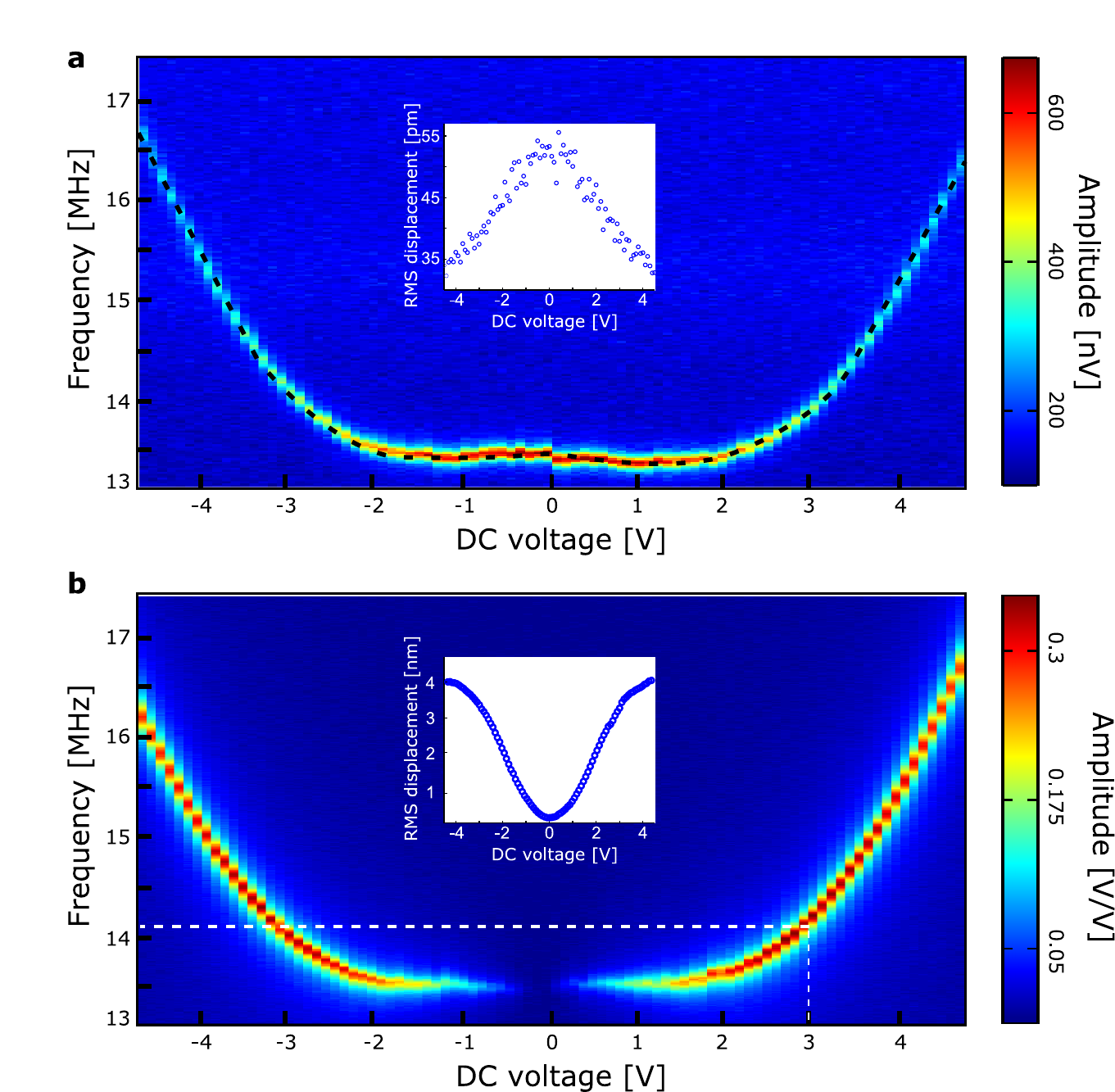}
	\caption[width=0.5\textwidth,justification = justified]{\textbf{Electrostatic tuning of the fundamental mode.} (a) Frequency response of the Brownian motion (color scale) as a function of the applied dc-voltage. The black dashed line indicates the resonance frequency. Inset: RMS amplitude of the thermal motion, showing an optimum at $\mathrm{V_{dc}}$ = 0. (b) Frequency response of the driven motion as a function of applied DC voltage. The driven motion was mapped at $\mathrm{V_{dc}}$ = 3 V, at a resonance frequency of $\mathrm{f_1}$ = 14.1 MHz.}
\end{figure}

\subsection*{4.	Peak-force Atomic Force Microscopy of the nanodrum surface}

To investigate the origin of the distortion of the higher order mode shapes, the local stiffness of the drum was mapped using Peak-Force Atomic force Microscopy (PF-AFM). Figure S4a shows the maximum displacement at each point of the drum using 5 nN as a setpoint for the PF-AFM. It reveals a wrinkle-like feature, which is indicated by the black dashed line. From the force-displacement curves, the effective spring constant at every point is extracted and it is shown in Figure S4b. It reveals that the wrinkle manifests as a source of uniaxial tension along its direction, denoted by $\mathrm{F_T}$.
To get a qualitative image of how this feature affects the mode shapes, we simulated the mode shapes using COMSOL®, in which the location of the wrinkle is included by imposing a clamped boundary condition. Figure S4c shows the simulated mode shapes of the measured modes. Clearly, the wrinkle not only breaks the degeneracy of modes (1,1), (2,1), and (1,2), but it also defines the direction of the nodal lines for the higher mode shapes. Taking the feature into account, the simulated mode shapes bear better resemblance with the measured ones, and they explain the unusual shapes of mode (0,2), which loses its radial symmetry, and mode (2,1) which shows merging of two antinodes along the diameter of the drum.

\begin{figure}[h]
	\includegraphics[width=0.5\textwidth]{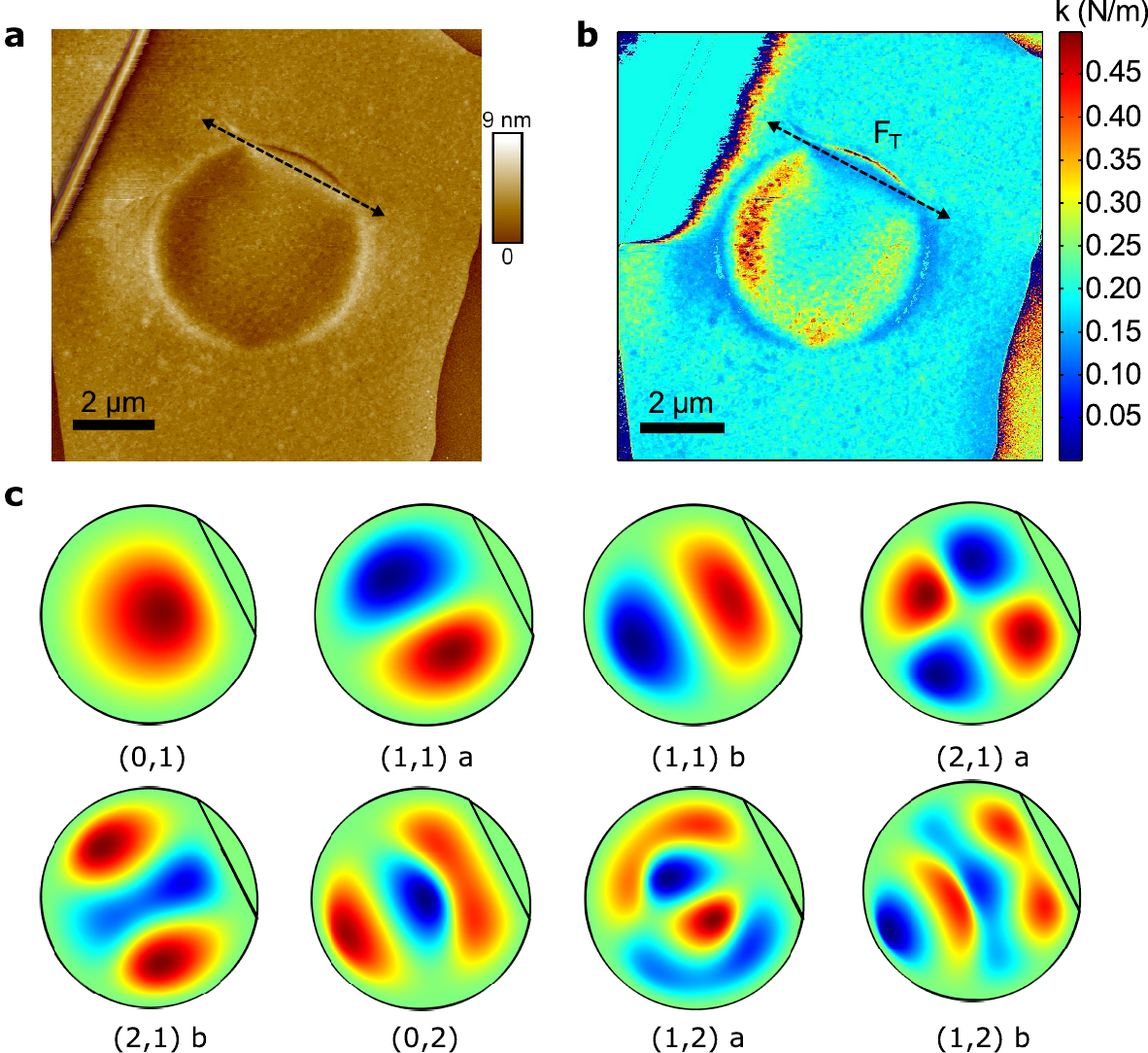}
	\caption[width=0.5\textwidth,justification = justified]{\textbf{Local stiffness of the drum taken by PF-AFM.} (a) Maximum displacement at a peak force of 5 nN. The AFM scan reveals a wrinkle indicated by the dashed line. (b) Effective local linear spring constant, k, as calculated by taking the derivative of the force with respect to deflection for small deflections. (c) Simulated mode shapes for the measured modes for a drum assuming a tensile force FT along the dashed line indicated in (a) included in the model as a fixed edge. Simulated mode shapes are rotated with respect to the AFM images to match the orientation of the sample while the mode shape imaging was performed.}
\end{figure}

\subsection*{5.	Measuring other parameters with spatial resolution}
Besides the height of the resonance peak, other characteristics can be plot as a function of the position on the drum. For each mode and at each position, one can for instance plot the noise floor, the resonance line-width, or derived parameters such as the ratio between the resonance frequencies. Of particular interest is the spatial distribution of the resonance frequencies, as the drum is heated by the probe laser. Figure S5a shows a spatial map of resonance frequency of the driven fundamental mode, $\mathrm{f_1}$. Clearly, the resonance frequency varies over the drum, and it is lowest at the circumference and maximizes in the centre. This is the result of heating of the system by the laser: as the reflectivity of the Si is lower than that of the AuPd, the sample heats up slightly more when probing the motion at the drum centre. The thermal expansion of the substrate induces additional tension in the membrane, which increases its resonance frequency. The induced tension, and thus the resonance frequency of the drum, therefore maximizes when the laser hits the centre of the drum. A map of the optical reflectivity R of the device, shown in Figure S5b, confirms the increased optical absorbance when measuring at the center of the drum.\\

\begin{figure}[h]
	\includegraphics[width=0.5\textwidth]{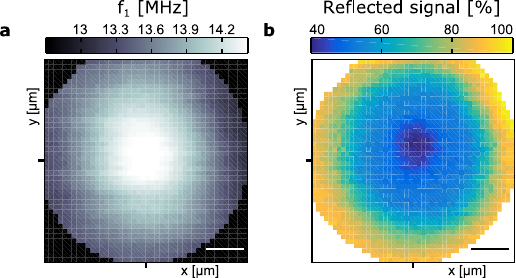}
	\caption[width=0.5\textwidth,justification = justified]{\textbf{Spatial map of the resonance frequency of the fundamental mode.} (a) Due to the different effective heating across the drum, its resonance frequency varies by 7 \%. The scale bar is 1 μm. (b) Reflectivity of the device, as represented by the dc output of the photodiode. The plotted signal, which is normalized to the reflectivity at the AuPd, clearly shows the increased absorption at the centre of the drum. }
\end{figure}

\subsection*{References}
[S1]	Blake, P. et al. Making graphene visible. Appl Phys Lett 91, 063124 (2007)

\begin{thebibliography}{10}
\bibitem{mode31}
\bibinfo{title}{The motion of mode (3,1), with an expected resonance at
  $\mathrm{37.4\,MHz}$, is not detected. {A} possible explanation is that the
  nodal lines in this device are spaced close to the diffraction limit,
  resulting in a low displacement contrast.}
\end{thebibliography}
\end{document}